# Verification of Fast Ion Effects on Turbulence through Comparison of GENE and CGYRO with L-mode Plasmas in KSTAR

**Donguk Kim[1#], Taeuk Moon[2#], Choongki Sung[1*], Eisung Yoon[2**], Sumin Yi[3], Jisung Kang[3], Jae-Min Kwon[3], Tobias Görler[4], Emily Belli[5], and Jeff Candy[5]**

[1] Department of Nuclear & Quantum Engineering One, Korea Advanced Institute of Science and Technology, Daejeon, South Korea
[2] Department of Nuclear Engineering, Ulsan National Institute of Science and Technology, Ulsan, South Korea
[3] Korea Institute of Fusion Energy, Daejeon, South Korea
[4] Max Planck Institute of Plasma Physics, Garching, Germany
[5] General Atomics, San Diego, California, United States of America

[#] Both are the 1st authors as they equally contribute.
*E-mail: choongkisung@kaist.ac.kr
**E-mail: esyoon@unist.ac.kr

### Abstract

This study presents a cross-verification of fast ion effects on turbulence through a systematic comparison of two leading gyrokinetic codes, GENE [F. Jenko et al., *Phys. Plasmas* **7** 1904-1910 (2000)] and CGYRO [J. Candy et al, *J. Comput. Phys.* **324** 73-93 (2016)], using L-mode plasma profiles from KSTAR for local linear and nonlinear electromagnetic simulations. The focus is on the impact of fast ions and rotation effects on energy flux, aiming to identify the similarities and differences between these codes in the context of turbulence transport research. The analysis shows consistency in linear stability results, fractional changes in energy flux, changes in the distribution of energy fluxes, fluctuations and phase angle with fast ions, and zonal shearing between the codes. However, discrepancies arise in absolute thermal energy levels and rotation effects on energy transport, especially in the presence of fast ions. The study underscores the critical importance of phase angle analysis in gyrokinetic code verification, particularly when assessing fast ion effects on turbulence. Additionally, it highlights the need to examine quantities at lower levels of the primacy hierarchy, as discrepancies at lower levels can lead to divergent results at higher levels. These findings indicate the necessity for further investigation into these discrepancies and the novel phase angle structures observed, contributing to the advancement of accurate transport predictions in fusion plasmas.

---

## 1. Introduction

Accurate prediction of transport phenomena in fusion plasmas is crucial for assessing the performance of prospective fusion devices as transport processes significantly influence plasma confinement and overall behavior. Therefore, a verification and validation (V&V) study of the current state-of-the-art transport models constitutes a prerequisite initial step prior to their extensive utilization in developing valid transport models. It is widely recognized that this transport is driven by turbulence originated from drift wave instabilities, for which gyrokinetic theory has been successfully employed over the past three decades to describe and predict the turbulent transport [1–4]. Consequently, extensive V&V studies of gyrokinetic codes have been conducted in various physics topics and operation modes [5–29].

Effects of fast ions generated by fusion reactions and external heating are anticipated to be significant in future fusion plasmas. Therefore, research on fast ion physics has emerged as a crucial focus area in nuclear fusion plasma studies [30,31]. While these energetic ions are essential for plasma heating and sustainment, they can also potentially drive magnetohydrodynamic (MHD) instabilities [32–34,30]. Furthermore, fast ions can influence turbulence and transport through mechanisms such as dilution and interaction between MHD modes and turbulence [35–41]. These effects have been studied actively using various gyrokinetic codes [27,35–40]. However, despite the necessity for verification as the prerequisite step prior to exploring the fast ion transport physics, a comprehensive cross-verification exercise involving gyrokinetic code predictions has not been undertaken thus far, to the best of our knowledge. This lack of verification motivates the current study, which aims to compare simulation results for a discharge with fast ions from two preeminent gyrokinetic continuum codes widely employed in fusion community, CGYRO [42] and GENE [43,44].

Verification of gyrokinetic codes has been carried out in various aspects in the nuclear fusion community. As zonal flow is an important player in microturbulence simulation, Rosenbluth-Hinton residual flow tests [45,46] become a standard procedure of verification for newly developing gyrokinetic codes [47–53], which is one of the well-known examples of comparison between analytic theory and numerical code results. In addition, significant efforts on cross-verification have been reported among different combinations of gyrokinetic codes [54,10,55] for various comparison conditions and physics phenomena. In the perspective of verifications, linear growth rate, frequency and ballooning mode structure for linear simulations and power spectrum and level of transport for nonlinear simulation are investigated under several interesting conditions such as gyrokinetic/gyrofluids [56,54], electrostatic/electromagnetic simulation [57,55,58], adiabatic response/kinetic electron [55], collisionless/collisional simulation [53,59,60], analytic/realistic equilibrium [61], and Lagrangian/Semi-Lagrangian/Eulerian codes [52,62,18].

During our research, we became aware of Ref. [63], which suggests that local simulations incorporating fast ions require a larger radial domain compared to conventional local simulations particularly for unstable and, to some degree, marginally stable energetic particle modes. As noted in the Ref. [63], long potential structures extend radially to the edge of the domain, causing short-circuiting, a phenomenon we also observed. By increasing the radial domain to ~512 $\rho_s$, where $\rho_s$ is an ion gyroradius of acoustic speed, we were able to encompass the longest radial potential structures within the radial domain. However, this configuration exceeds the global radial domain size of the KSTAR reactor and undermines the primary advantage of local simulations, which is their computational efficiency. Consequently, adopting such a large domain is deemed impractical for the purpose of our study. Furthermore, the primary objective of our research is to verify the CGYRO and GENE codes. Therefore, we have continued to focus on local simulations, maintaining domain sizes that ensure both the feasibility and relevance of our findings. This approach aligns with the overarching goal of our study, which is the verification and comparison of these simulation tools within the confines of local simulation parameters.

In this study, we conducted linear and nonlinear electromagnetic simulations for fully stripped Deuterium and Carbon ions against an L-mode KSTAR discharge for cross-verification in a reconstructed magnetic equilibrium [26]. Considering the characteristics of the selected discharge, the role of fast ions was deemed crucial. Consequently, the simulation cases were categorized based on the presence or absence of Deuterium fast Ions and the consideration of plasma flow and its shear in the context of ablation tests. The detailed input parameters for the simulation setups are elucidated in Section 2. Following the prescribed methodology applied in this verification study, Section 3 presents the linear simulation results, encompassing the growth rates and real frequencies of the most unstable mode. Furthermore, the energy fluxes obtained from the nonlinear simulations are scrutinized through spectrum analysis to elucidate underlying characteristic physics and ensure consistent behavior between the codes. Finally, Section 4 summarizes the verification outcomes and their implications while discussing novel findings uncovered during this verification study.

## 2. Verification Setup

For cross-verification of the electromagnetic linear and nonlinear local gyrokinetic simulations by CGYRO and GENE, the KSTAR L-mode discharge 21631 with NBI (neutral beam injection) heating was used as a common reference. This discharge has been utilized for the validation study in KSTAR [26]. Major global parameters of this discharge are magnetic field on axis $B_T = 2.5$T, plasma current $I_P = 0.6$MA on the flat top, total NBI power $P_{NBI} = 2.9$MW, line average density $\bar{n}_e \sim 2\times10^{19}$m$^{-3}$, safety factor at 95% of a normalized poloidal flux $q_{95} \sim 5$, and normalized beta $\beta_N \sim 1.33$. Effective charge $Z_{eff} (=\Sigma_j Z_j^2 n_j/n_e)$ was assumed to be 2.0 with a flat profile in the present analysis, where j is over ion species.

In addition, MHD effects are ignored as MHD activities are weak (~0.3G) in this discharge. More detailed information about this discharge can be found in Ref. [26].

Table 1. Input parameters used for gyrokinetic simulations. Here, experimental profiles in KSTAR (shot 21631, t=2050ms) were used as input parameters.

| Input parameter | Description | Values |
|---|---|---|
| $R_0$ | Major radius (meter) | 1.800 |
| $a$ | Minor radius (meter) | 0.485 |
| $r/a$ | Center of radial simulation domain | 0.5 |
| $B_0$ | Magnetic field strength at axis (T) | 2.492 |
| $q_0$ | Local safety factor | 1.895 |
| $\hat{s}$ | Magnetic shear | 0.444 |
| $n_e$ | Electron density ($m^{-3}$) | 2.52 |
| $n_i/n_e$ | Thermal Deuterium density | 0.8 (I)<br>0.684 (III) |
| $n_c/n_e$ | Carbon density | 0.033 |
| $n_{fi}/n_e$ | Fast Deuterium density | 0.116<br>(III/IV only) |
| $R/L_{ne}$ | Gradient scale length of electron density | 6.828 |
| $R/L_{ni}$ | Gradient scale length of thermal Deuterium density | 6.828 (I/II)<br>5.713 (III/IV) |
| $R/L_{nc}$ | Gradient scale length of Carbon density | 6.828 |
| $R/L_{nfi}$ | Gradient scale length of fast Deuterium density | 13.37<br>(III/IV only) |
| $T_e$ | Electron temperature (KeV) | 1.49 |
| $T_i/T_e$ | Thermal Deuterium temperature | 1.124 |
| $T_c/T_e$ | Carbon temperature | 1.171 |
| $T_{fi}/T_e$ | Fast Deuterium temperature | 15.88 |
| $R/L_{Te}$ | Gradient scale length of electron temperature | 8.568 |
| $R/L_{Ti}$ | Gradient scale length of thermal Deuterium temperature | 5.352 |
| $R/L_{Tc}$ | Gradient scale length of Carbon temperature | 5.826 |
| $R/L_{Tfi}$ | Gradient scale length of fast Deuterium temperature | 1.478<br>(III/IV only) |
| $k_{y,min}$ | Minimum $k_y$ value (product base) | 0.067 |

Input profiles and equilibrium parameters were generated through the iteration process among profile analysis, transport analysis, and equilibrium reconstruction. Profiles including electron density and temperature, carbon temperature, and toroidal velocity were obtained through polynomial fitting based on the raw data from diagnostics represented in Ref. [26]. Density and temperature profiles of the main thermal ion and fast ion were calculated from the transport analysis code, TRANSP [64] coupled with NUBEAM [65], which provides fast ion information generated by NBI through the Monte Carlo technique. Equilibrium was reconstructed by an equilibrium code, EFIT [66] with the constraints of total pressure profiles, calculated from profile analysis and transport analysis, and magnetic field information obtained from magnetic diagnostics [67] and motional Stark effect (MSE) diagnostic [68]. Furthermore, the determination of the poloidal velocity necessary for the computation of the ExB shearing rate, a factor impacting turbulence dynamics, was conducted by NEO [69], relying upon neoclassical theoretical frameworks.

Table 2. Ablation setting for the 4 cases for gyrokinetic verification. O and X denote the inclusion and exclusion, respectively

| **Case number** | **Ablation setting for nonlinear simulations** | |
|---|---|---|
| | **With fast Deuterium** | **With flow & flow shear** |
| I | X | X |
| II | X | O |
| III | O | X |
| IV | O | O |

The derived information including the equilibrium and plasma parameters at the position of $r/a$ = 0.5 presented in Ref. [26] was commonly used for input parameters of CGYRO and GENE. Both codes used the Generalized Miller geometry [70,71] to represent an equilibrium magnetic field based on gEQDSK from EFIT. Considering the beta is not considerably big ($\beta$~0.005), we ignored $\delta B_\parallel$ while $\delta A_\parallel$ is kept. Nonlinear gyrokinetic equations for deuterium ion ($D^+$), fully stripped carbon ($C^{6+}$), and electrons ($e^-$) were locally solved with both codes, and the density ($n_c$) and density gradient ($a/L_{nc}$) of carbon were calculated using assumed $Z_{eff}$ = 2.0 and the quasi-neutrality relation. The Sugama collision operator [72] was used in both codes with automatically calculated collisionality for the given local density and temperature. In this study, we conducted ablation tests on four cases, as presented in Table 2, to comparatively verify CGYRO and GENE, two codes designed for evaluating turbulence-driven transport in nuclear fusion reactors. For the "With flow & flow shear" option, we toggled three flow-related physical quantities together (toroidal flow, toroidal flow shear, and ExB shear) that are associated with turbulent suppression in tokamaks, as well as the presence of fast ions, which are considered another factor for turbulent suppression in recent studies [35,73,36–38,74,75,39–41,76–79]. Maxwellian distribution function is utilized for the initial distribution function of the fast ions as well as the other species in this work. Note that all results presented in this article are obtained from specific versions of GENE 2.0 (Commit: cd8f7485) and CGYRO (Commit: aa4f702).

For the tests, the time resolution in both codes used the option to automatically calculate timestep based on the eigenvalue. The spatial resolution using straight field-aligned coordinates is decided by a convergence test, which yielded ($n_{x0}$, $n_{ky0}$, $n_{z0}$, $n_{v0}$, $n_{w0}$) = (256, 16, 64, 32, 16) for GENE, where $n_{x0}$, $n_{ky0}$, $n_{z0}$, $n_{v0}$, and $n_{w0}$ are the number of grid points in radial direction, bi-normal Fourier modes, parallel direction to magneitc field, parallel velocity direction, and magnetic moment direction, respectively, and $(n_{radial}, n_{\xi}, n_{energy}, n_{\theta}, n_{\varphi})$ = (216, 24, 8, 56, 16) for CGYRO, where $n_{radial}, n_{\xi}, n_{energy}, n_{\theta}$ , and $n_{\varphi}$ are the resolution on parameters for radial, pitch angle, energy, poloidal, and binormal grids, that were used throughout nonlinear simulations of this study. Average and uncertainty of flux were obtained by calculating the average and standard deviation of the saturated phase more than 100 $a/c_s$.

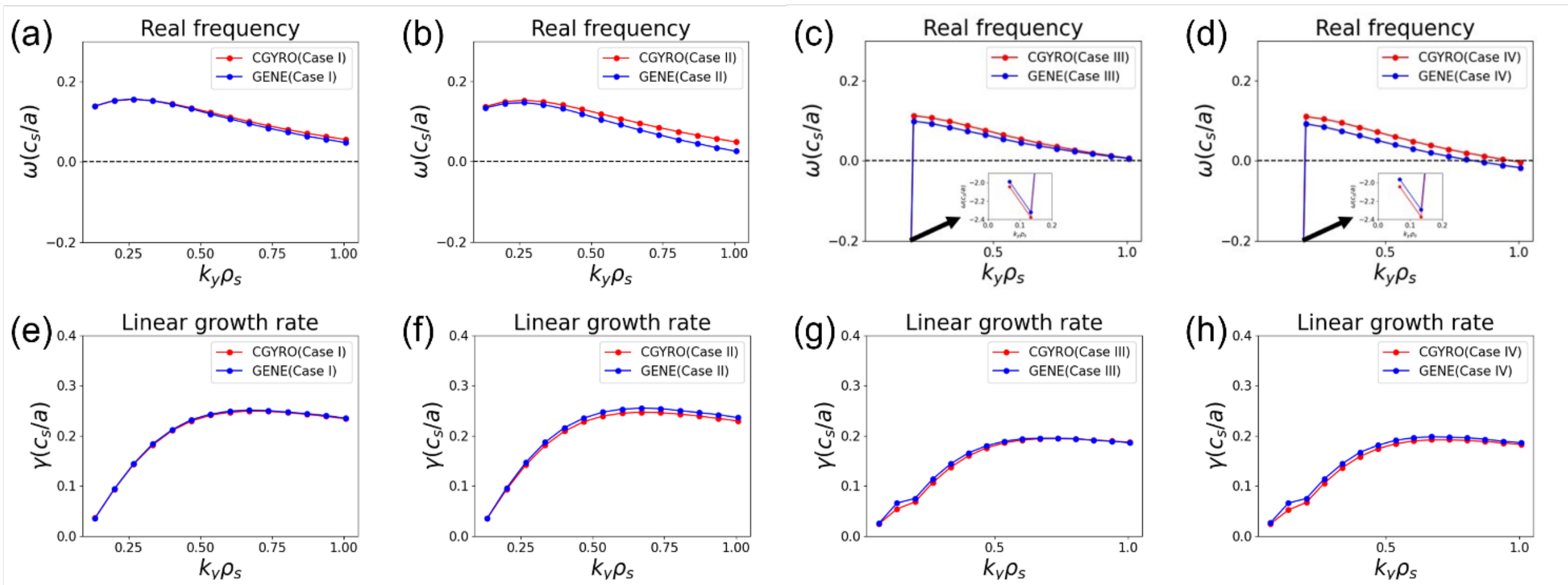

Figure 1. Real frequency (top) and linear growth rate (down) of the most unstable mode for Cases I-IV as a function of $\boldsymbol{k_y\rho_s}$. CGYRO and GENE results are shown as red and blue colors, respectively.

## 3. Cross-comparison of the Simulation Results

### *3.1 Linear Stability Analysis*

The real frequencies and linear growth rates for each $k_y\rho_s$ mode from cases I to IV are examined as shown in Figure 1, where $k_y$ and $\rho_s$ are binormal wave number and gyro radius of reference acoustic speed, respectively. Red color denotes the CGYRO results while blue is for GENE. The (+) and (-) signs of real frequencies are electron and ion diamagnetic directions, respectively. ExB shearing rate was set to zero when rotation effects were considered in the linear stability analysis performed here. Notably, experimental values are used for the other rotation relevant parameters, such as Mach number and parallel shearing rate, in the cases taking account of rotation effects, i.e., Cases II and IV.

Both gyrokinetic codes predict that the most dominant mode is in the electron diamagnetic direction for all cases. The real frequency and linear growth rate for the case I show good agreement. Linear results of the two gyrokinetic codes are still in good agreement in the case II, where the rotation effects were included. The good agreement in linear simulation results despite the inclusion of the rotation effect is consistent with previous cross-verification study results [21] using GYRO and GENE . In case III where fast ions were included, linear growth rate decreased in both codes, which is consistent with linear simulation results in previous studies [35,37,38]. Furthermore, cases involving fast ions (cases III and IV) exhibit a discontinuous jump in real frequency at a relatively long wavelength, $k_y\rho_s{\sim}0.2$, which moves towards the ion diamagnetic direction. This observed discontinuity in both CGYRO and GENE simulations is associated with the emergence of long wavelength modes in $k_y\rho_s < 0.2$, likely due to the presence of fast ions. In the cases III and IV, relative discrepancy between the two codes in the real frequency of the observed long wavelength mode is ~3% while the discrepancy of linear growth rate is at most ~33% between CGYRO and GENE. Although the difference in the linear growth rate between the two codes is appeared to be significant in terms of fractional change, the absolute difference is less than 0.014 $c_s/a$, much smaller than the maximum linear growth rate observed in this analysis (~0.20 $c_s/a$). Overall, real frequency shows good agreement regardless of the inclusion of rotation effects and fast ions while the inclusion of fast ion species can increase the discrepancy in linear growth rate of long wavelength mode although its difference is still not remarkable in the absolute level.

### *3.2 Energy Transport Prediction*

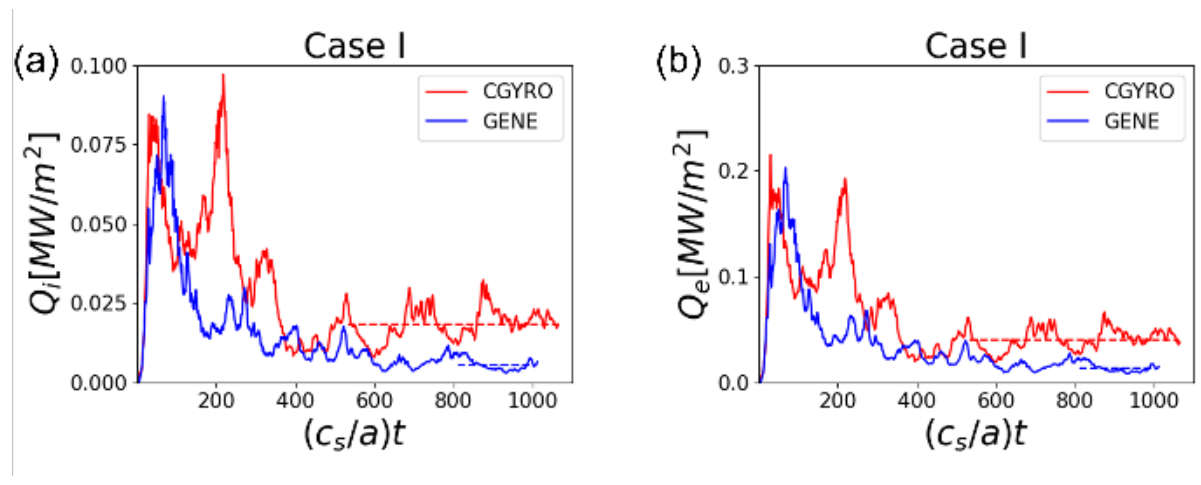

Figure 2. Temporal evolution of (a) ion and (b) electron energy flux levels predicted by nonlinear simulations for Case

I. Here, red and blue colors denote the CGYRO and GENE results, respectively.

Next, we compared total thermal energy flux (i.e., sum of electrostatic and electromagnetic contributions) levels predicted by nonlinear simulations run by the two codes. Figures 2(a) and (b) show the temporal evolution of ion and electron energy flux levels for Case I, respectively. Here, thermal ion energy flux includes the fluxes from both main ion species and impurity, which corresponds to deuterium and carbon, respectively, in this study. Nonlinear runs generally have two phases, initial linear phase and nonlinearly saturated phase. Typically, instabilities are first developed in the initial linear phase, as predicted by linear stability analysis. Afterward, these instabilities are regulated through nonlinear effects, such as zonal shearing, finally reaching to the saturated phase. Since temporal evolution in the initial linear phase can be different due to numerical setup in each code, such as initial conditions, we utilized the energy flux levels averaged during the saturated phase, presented as dashed line in Figure 2, for the quantitative comparison between CGYRO and GENE in this study.

The results show that CGYRO predicts energy flux levels for both thermal ions and electrons to be approximately three times higher than those predicted by GENE. It is important to note that Case I does not include either fast ions or rotation effects, indicating that the discrepancy is not due to these factors. Therefore, while clarifying the quantitative discrepancy of Case I is beyond the scope of this study and should be addressed in future research, we investigate the effects of fast ions and rotation on gyrokinetic predictions by examining the relative changes in the simulated quantities including energy flux, as each effect is included in other test cases (Cases II – IV).

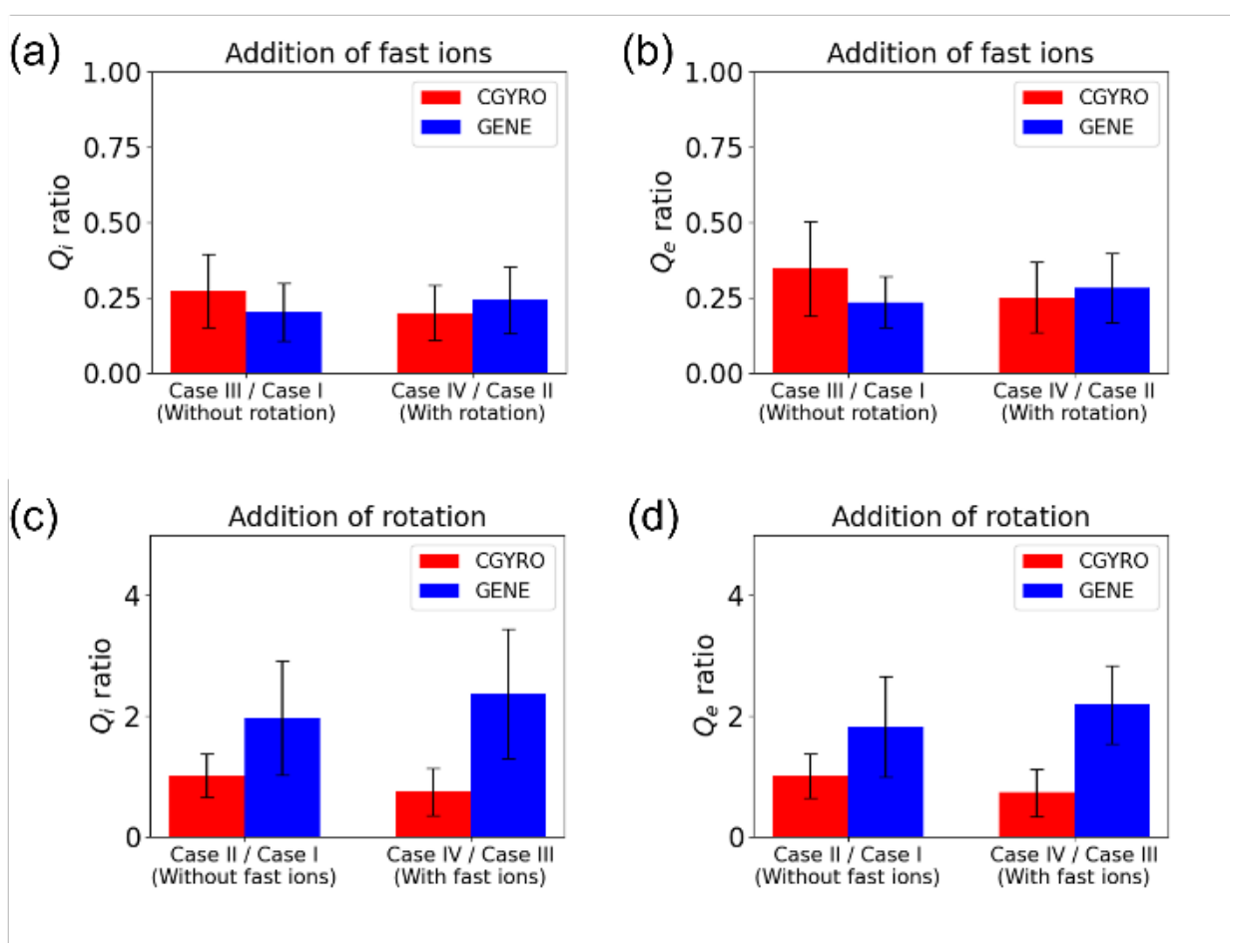

Figure 3. Fractional changes in the energy flux levels as fast ions (top) or rotation effects (bottom) are incorporated.

Figure 3 illustrates the fractional change in the energy flux levels as fast ions or rotation effects are incorporated into the simulation. In Figures 3(a) and (b), both CGYRO and GENE predict a significant reduction in energy flux with the addition of fast ions. Without considering rotation effects, CGYRO predicts reductions of approximately 73% for thermal ion energy flux and 65% for electron energy flux upon the addition of fast ions. Similarly, GENE predicts reductions of around 80% for thermal ion energy flux and 77% for electron energy flux, closely aligning with CGYRO results both qualitatively and quantitatively within their uncertainties. We observed closer agreements in fractional change of energy flux levels when rotation effects were present than in its absence. In this case, both codes predicted approximately 75-80% reduction of both thermal ion and electron energy flux levels.

However, discrepancies arise in the fractional change due to consideration of rotation effects as observed in Figures 3 (c) and (d). When rotation effects were added without fast ions, i.e., from Case I to Case II, CGYRO predicts almost identical thermal energy flux levels within 1%, while GENE predicts 80-90% increase in thermal energy flux levels. Although the discrepancy lies within their uncertainties, it is not negligibly small. This result is consistent with the previous results showing discrepancy between GYRO and GENE when rotation effects were considered without fast ions [21]. Larger discrepancy was observed when fast ions were added. CGYRO predicts approximately 25-30% reduction in thermal energy flux levels as rotation effects were added with fast ions, i.e., from Case III to Case IV. In contrast, we observed approximately two times higher thermal energy flux levels in Case IV compared to Case III from GENE simulation results. Overall, inclusion of rotation effects result in 80-100% increase in thermal energy flux levels in GENE simulations while CGYRO predicts the opposite trend, similar or decreased thermal energy flux levels, by 25-30% in the identical case.

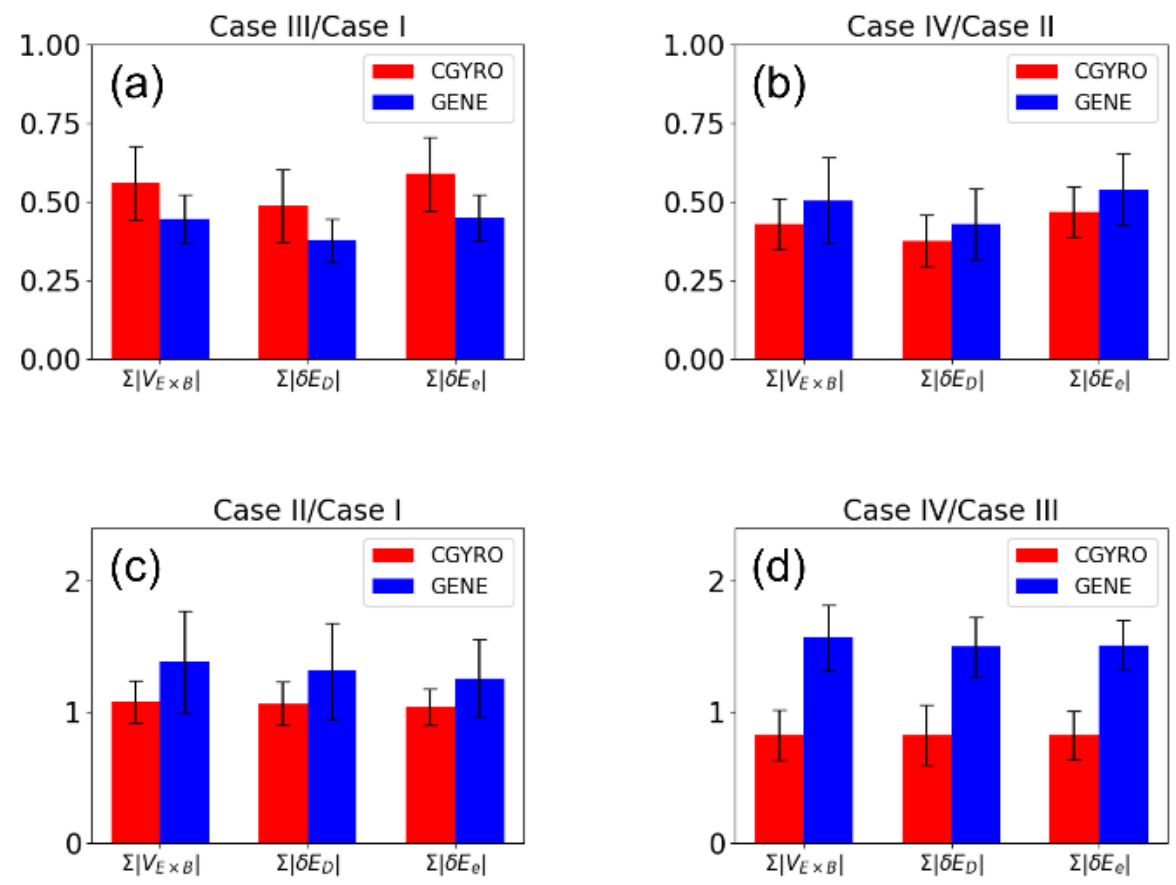

Figure 4. Comparison of the sum of magnitudes of the fluctuating components including $\widetilde{\boldsymbol{V}}_{\boldsymbol{E\times B}}$ and $\widetilde{\boldsymbol{E}}$ in $\boldsymbol{-0.51 \le k_x\rho_s \le 0.51}$ and $\boldsymbol{0.2 \le k_y\rho_s \le 0.6}$ under different conditions: the presence or absence of fast ions and rotations. $\tilde{\boldsymbol{V}}_{\boldsymbol{E}\times\boldsymbol{B}}$ and $\tilde{\boldsymbol{E}}$ denote fluctuation of $\mathbf{E}\times\mathbf{B}$ flow velocity and energy, respectively.

To facilitate a more detailed comparative analysis between the codes, we conducted an investigation into the individual contributions of the fluctuating components to the turbulence-driven energy flux. Since our simulation results indicate that the majority of energy flux is generated by electrostatic turbulence (≥ 97%) we extracted the fluctuating components $\tilde{V}_{E\times B}$ and $\tilde{E}$ from the simulation data referring to following equation of electrostatic energy flux.

$$\tilde{Q}_j = \mathrm{Re}\left[\tilde{V}_{E\times B}\tilde{E}_j^*\right] = \left|\tilde{V}_{E\times B}\right|\left|\tilde{E}_j\right|\cos\alpha_{V_{E\times B}E_j}. \qquad (1)$$

Here, $\tilde{Q}_j$ is electrostatic part of the energy flux driven by turbulence for species $j$ and $\alpha_{V_{E\times B}E_j}$ is phase angle difference between the E×B veloctiy fluctuation $\tilde{V}_{E\times B}$ and energy fluctuation $\tilde{E}_j$ . Figure 4 presents a comparison of the magnitudes of the fluctuating components summed in k-region where mainly contributed in $-0.51 \leq k_x\rho_s \leq 0.51$ and $0.2 \leq k_y\rho_s \leq 0.6$ across the *k*-spectra (which is found in Figure 5) under different conditions: the presence or absence of fast ions (Figures 4(a) & (b)) and the presence or absence of rotation (Figures 4(c) & (d)).

From Figures 4(a) and (b), we observe that the presence of fast ions leads to a qualitative reduction in the magnitude of E x B fluctuations and in the energy fluctuations of deuterium and electrons regardless of rotation effect in both codes. Quantitatively, in the absence of rotation, the reduction in fluctuations is a little bit more pronounced in GENE, while CGYRO exhibits a similar or slightly greater reduction when rotation is included. Overall, the changes in fluctuations remain at a comparable level within the error bars or slightly differ.

Examining the impact of rotation on the magnitude of the fluctuating quantities, as shown in Figures 4(c) and (d), we find that in the absence of fast ions (Figure 4(c)), CGYRO shows a slight increase in fluctuation magnitude when rotation is present, while GENE predicts larger levels of change, 25-40%, although the difference between two codes is within the error bars. Moreover, when fast ions are present (Figure 4(d)), CGYRO exhibits approximately 20% decrease in fluctuation magnitudes with rotation, whereas GENE shows approximately 40-50% change in fluctuation amplitudes. This finding aligns closely with the observed increase in the energy flux ratio shown in Figure 3.

While the sum of fluctuation magnitudes related to energy flux changes may suffice for a basic level of verification, it has limitations to fully elucidate the underlying mechanisms. Rigorous verification at a more fundamental level is required for a comprehensive understanding. Therefore, we also investigated the relevant fluctuating quantities, i.e., $\tilde{V}_{E\times B}$ and $\tilde{E}$, and the phase angle between them in $k_x$ and $k_y$ space between CGYRO and GENE for the distribution of thermal energy flux, where $k_x$ and $k_y$ are the wave numbers at normal to the flux surface and binormal directions, respectively. To preface the following discussion, our detailed analysis reveals that the phase angle, representing the coherence of fluctuation quantities in electrostatic energy flux, plays a significant role in simulations, particularly when fast ions are present. This finding underscores the importance of considering the phase angle in verification processes, which was also demonstrated in previous studies [6,19], but note that this study extends its importance in gyrokinetic prediction of fast ion effects on turbulence.

Figures 5(a)-(d) display contours of the electrostatic component of electron energy flux, i.e., $Re[\tilde{V}_{E\times B}\tilde{E}_e^*]$, $\tilde{V}_{E\times B}$, $\tilde{E}_e$, and cosine of the phase angle between $\tilde{V}_{E\times B}$ and $\tilde{E}_e$ in Case I predicted by CGYRO, respectively. All contours are normalized by their respective maximum value to emphasize the distrbituion. The electron energy flux is primarily concentrated in the region where $-0.51 \leq k_x\rho_s \leq 0.51$ and $0.2 \leq k_y\rho_s \leq 0.6$ . Both $\tilde{V}_{E\times B}$ and $\tilde{E}_e$ also peak near this region, and the cosine of their phase angle approaches 1.0 in the same area. There are noticible activities from $k_y\rho_s > 0.6$ in the $\tilde{V}_{E\times B}$ contour and from $k_y\rho_s < 0.2$ in $\tilde{E}_e$ contour, but these activities do not contribute to the electrostatic electron energy flux since the energy flux is a result of the convolution of these quantities. A similar trend is observed in the contours of thermal ion species in Case I, although it is not shown here.

Figures 5(e)-(h) show the distributions of electrostatic electron energy flux and its fluctuating quantities for Case III, the case with fast ions but without rotation effects, predicted by CGYRO. Although the amplitude of energy flux is significantly reduced in the presence of fast ions as observed in Figures 3(a) and (b), the distribution of electrostatic electron energy flux still peaks in a similar region, $-0.51 \leq k_x\rho_s \leq 0.51$ and $0.2 \leq k_y\rho_s \leq 0.6$ , as in Case I. However, the distributions of fluctuating quantities in this case, shown in Figures 5(f) and (g), differs significantly from those in Case I. In both $\tilde{V}_{E\times B}$ and $\tilde{E}_e$ , the long-wavelength region where $k_y\rho_s < 0.2$ is the main contributor. However, this is not reflected in the energy flux contour because the cosine of the phase angle in this region is near zero, as shown in Figure 5(h). From the changes shown in Figures 4 and 5 by fast ions, the reduction in thermal energy flux with the addition of fast ions can be attributed to two main factors: the reduced amplitude of fluctuating quantities in the region where energy flux peaks without fast ions, i.e., $-0.51 \leq k_x\rho_s \leq 0.51$ and $0.2 \leq k_y\rho_s \leq 0.6$, and the out of phase relationship between $\tilde{V}_{E\times B}$ and $\tilde{E}_e$ in the long-wavelength region despite the appearance of long wavelength mode with fast ions.

Similar trend was observed in the GENE simulations as well. Figure 6 shows the distribution of electrostatic electron energy flux and its fluctuating quantities in Cases I and III predicted by GENE. In Case I, all distributions shown in Figures 6(a)-(c) peak in the region of $-0.51 \leq k_x\rho_s \leq 0.51$ and $0.2 \leq k_y\rho_s \leq 0.6$, similar to the CGYRO results. GENE also predicts a shift of the peak to the region of $k_y\rho_s < 0.2$ in the $\tilde{V}_{E\times B}$ and $\tilde{E}_e$ contours with the addition of fast ions, but this shift is not reflected in the energy flux contour due to the near-zero value of the cosine of the phase ange between $\tilde{V}_{E\times B}$ and $\tilde{E}_e$. A similar trend observed in the distribution of electrostatic energy flux for thermal ion species predicted by GENE as well. Notably, similar fast ion effects are observed in cases that consider rotation effects, i.e., from Case II to IV. This indicates that both codes consistently predict the impact of fast ions on changes in the distribution of electrostatic thermal energy flux and its fluctuating quantities. It is also noteworthy that both CGYRO and GENE predict a symmetric distribution of the phase angle with respect to $k_x$ and broader distribution with fast ions, as shown in Figures 5(d) and (h) and Figure 6(d) and (h).

To investigate the changes in the distributions of energy flux and its fluctuating quantities due to rotation effects in the presence of fast ions, we compared the distributions in Cases III and IV predicted by CGYRO and GENE, as shown in Figures 7 and 8. Both codes predict no significant change in the distributions when rotation effects are considered. This result supports the consistency between the two codes in predicting changes in the distribution of electrostatic energy flux and fluctuations when fast ions or rotation effects are introduced. It also suggests that the observed discrepancies in flux level changes due to rotational effects stem not from differences in the predicted distribution, but rather from variations in amplitude.

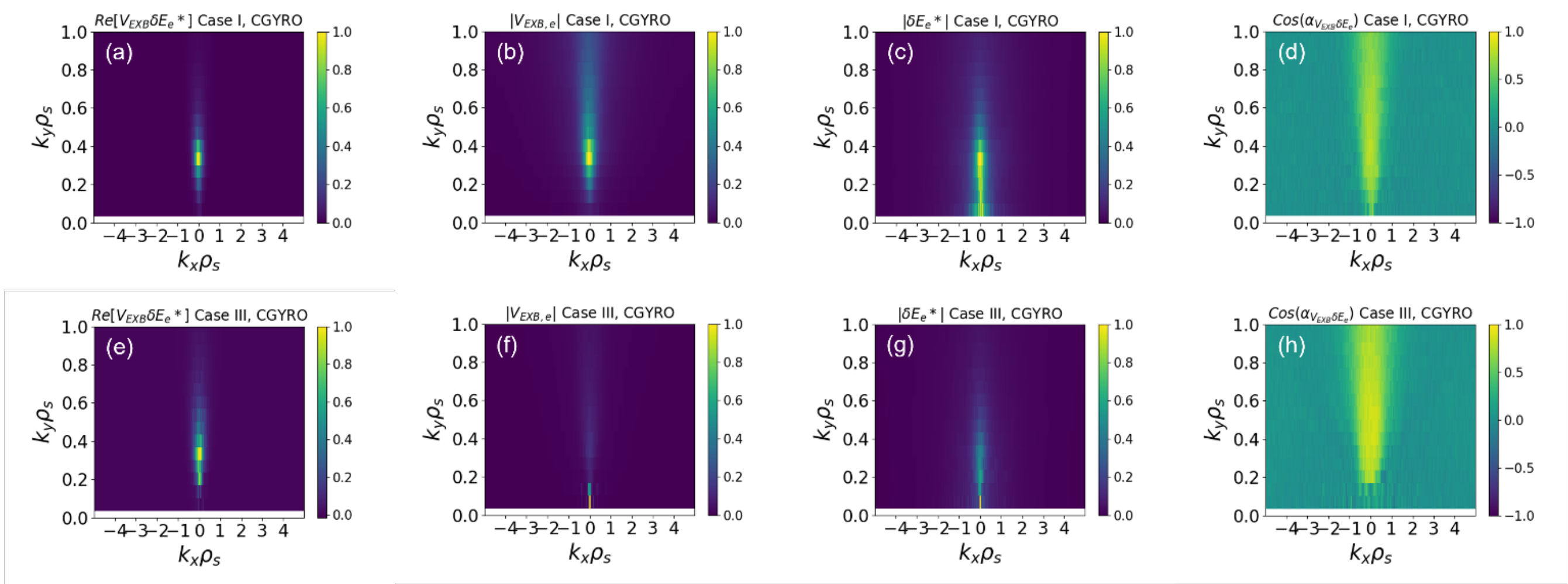

Figure 5. $\boldsymbol{k_x - k_y}$ contour of electrostatic electron energy flux ($\boldsymbol{Re[\tilde{V}_{E\times B}\tilde{E}_e^*]}$), and its fluctuating components ($\boldsymbol{\tilde{V}_{E\times B}}$, $\boldsymbol{\tilde{E}_e}$, and the cosine of phase angle between $\boldsymbol{\tilde{V}_{E\times B}}$ and $\boldsymbol{\tilde{E}_e}$) for Case I (top) and III (bottom) predicted by CGYRO

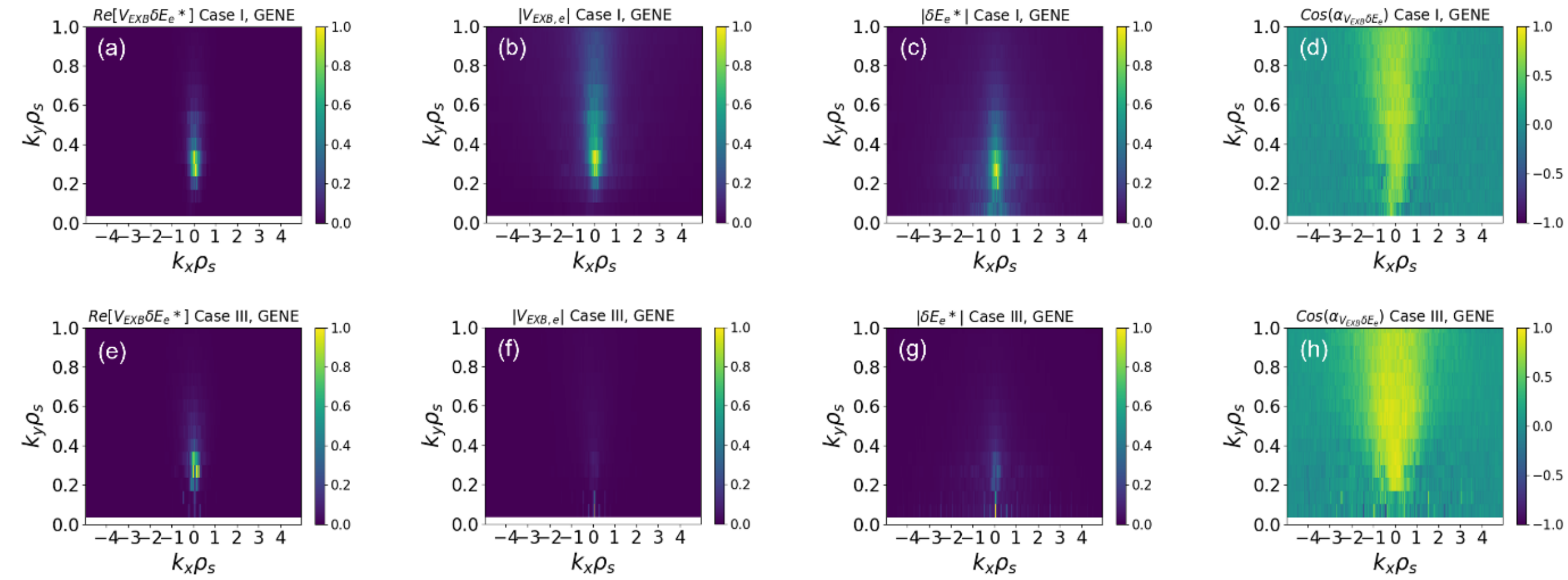

Figure 6. $\boldsymbol{k_x - k_y}$ contour of electrostatic electron energy flux ($\boldsymbol{Re[\tilde{V}_{E\times B}\tilde{E}_e^*]}$), and its fluctuating components ($\boldsymbol{\tilde{V}_{E\times B}}$, $\boldsymbol{\tilde{E}_e}$, and the cosine of phase angle between $\boldsymbol{\tilde{V}_{E\times B}}$ and $\boldsymbol{\tilde{E}_e}$) for Case I (top) and III (bottom) predicted by GENE

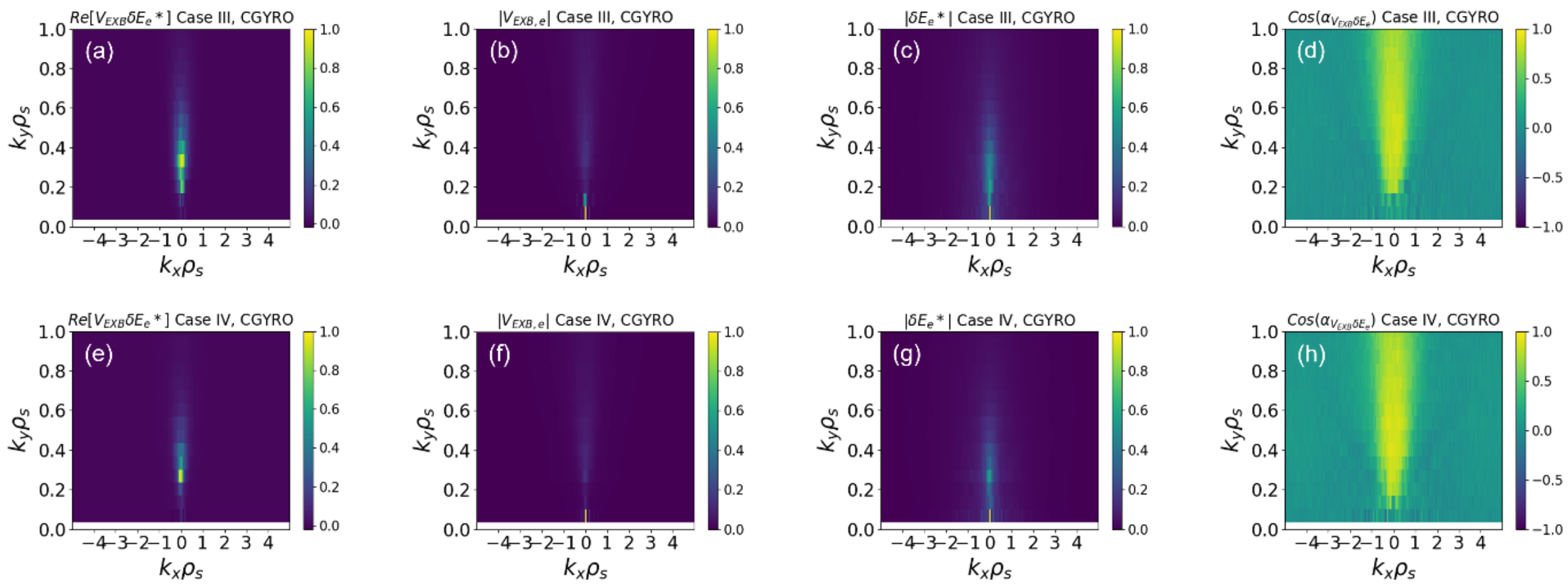

Figure 7. $\boldsymbol{k_x - k_y}$ contour of electrostatic electron energy flux ($\boldsymbol{Re}[\widetilde{\boldsymbol{V}}_{\boldsymbol{E\times B}}\widetilde{\boldsymbol{E}}_{\boldsymbol{e}}^{*}]$), and its fluctuating components ($\widetilde{\boldsymbol{V}}_{\boldsymbol{E\times B}}$, $\widetilde{\boldsymbol{E}}_{\boldsymbol{e}}$, and the cosine of phase angle between $\widetilde{\boldsymbol{V}}_{\boldsymbol{E\times B}}$ and $\widetilde{\boldsymbol{E}}_{\boldsymbol{e}}$) for Case III (top) and IV (bottom) predicted by CGYRO

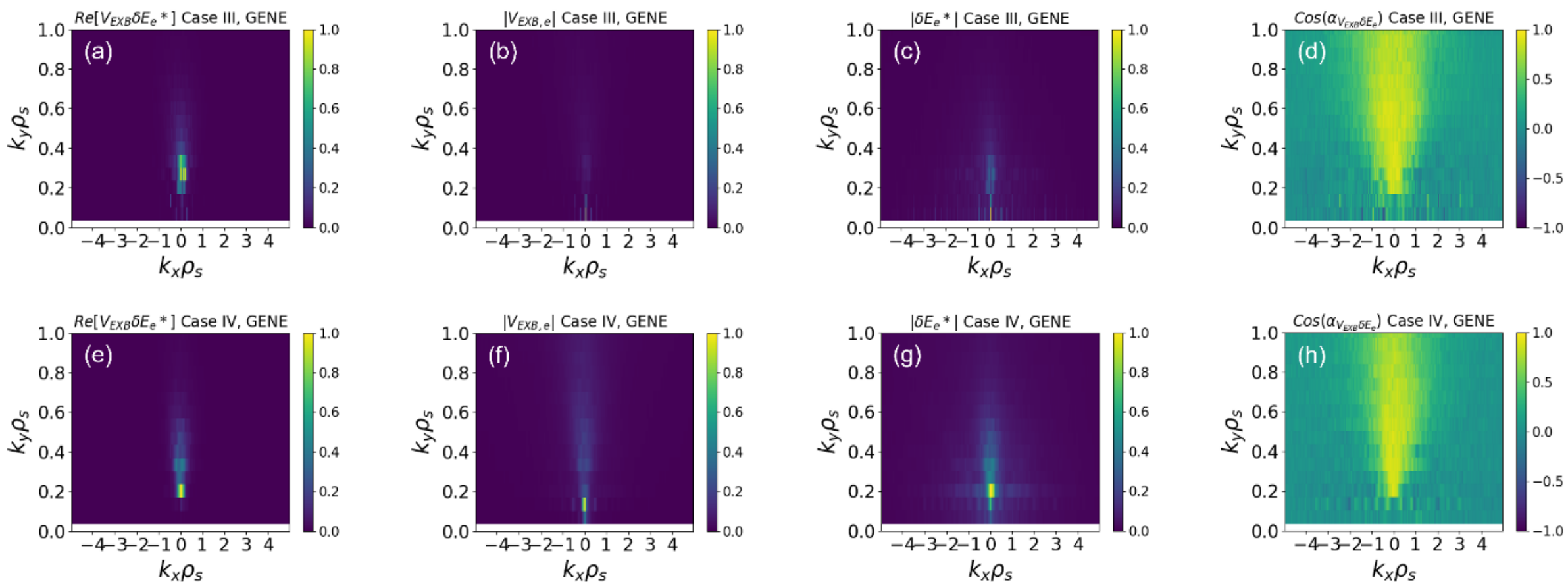

Figure 8. $\boldsymbol{k_x - k_y}$ contour of electrostatic electron energy flux ($\boldsymbol{Re}[\widetilde{\boldsymbol{V}}_{\boldsymbol{E\times B}}\widetilde{\boldsymbol{E}}_{\boldsymbol{e}}^{*}]$), and its fluctuating components ($\widetilde{\boldsymbol{V}}_{\boldsymbol{E\times B}}$, $\widetilde{\boldsymbol{E}}_{\boldsymbol{e}}$, and the cosine of phase angle between $\widetilde{\boldsymbol{V}}_{\boldsymbol{E\times B}}$ and $\widetilde{\boldsymbol{E}}_{\boldsymbol{e}}$) for Case III (top) and VI (bottom) predicted by GENE

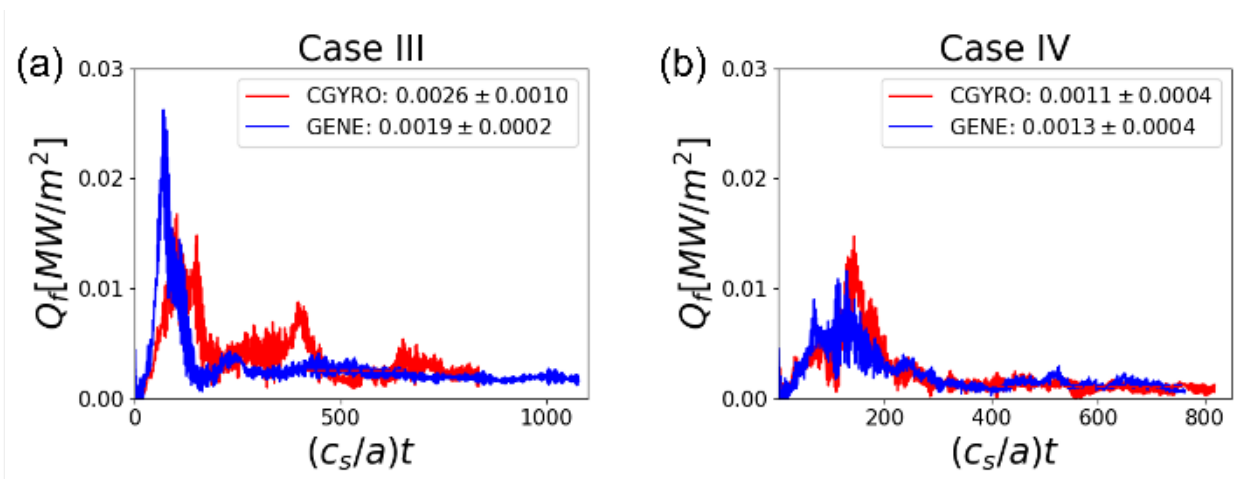

Figure 9. Temporal evolution of fast ion energy flux levels predicted by nonlinear simulations for (a) Case III and (b) IV.

### 3.3 Fast Ion Energy Transport Prediction

Figure 9 illustrates the temporal evolution of fast ion energy flux. While the two codes exhibit divergent time histories during the initial simulation phase, they converge within the standard deviation uncertainty after reaching saturation. Consequently, both codes demonstrate good agreement in the energy flux change ratio associated with the rotation effect in the presence of the fast ions. The agreement in the fast ion energy flux levels are consistent with the agreed fractional change in the thermal energy flux levels due to the addition of fast ions discussed in Section 3.1.

In line with the previous analyses, we have plotted the magnitudes and phase angle differences of the key factors in electrostatic energy transport in the $k_x$-$k_y$ space for fast ions, as shown in Figure 10. Notably, the dominant mode is appeared in the $k_y\rho_s$~0.2 region, aligns with the signature of fast ions identified in the linear analysis with fast ions (Figures 1(g) and (h)). It is also consistent with the changes in the thermal energy flux with the addtiion of fast ions. Specifically, compared to the absence of fast ions (CGYRO: Figures 5(b) and (c); GENE: Figures 6(b) and (c)), the dominant mode in $k_x$-$k_y$ space shifts to the lower $k_y$ region ($k_y\rho_s$~0.2) when fast ions are present (CGYRO: Figures 5(f) and (g); GENE:

Figures 6(f) and (g)). This indicates that the presence of fast ions induces a shift to the lower $k_y$ modes for the dominant modes, leading to a decrease in transport due to the phase misalignment between the two fluctuating quantities. The correlation between the presence of fast ions and the appearance of the dominant $k_y$ mode is illustrated in Figures 10(b) and (c) for CGYRO and Figures 10(f) and (g) for GENE. Another point of interest is the clear and distinct structures in the $k_x$-$k_y$ space phase angle differences shown in Figure 10(d) for CGYRO and Figure 10(h) for GENE. Both codes exhibit even parity of symmetry with respect to $k_x$. The interpretation of this phase angle symmetry is beyond the scope of the V&V objectives of this study and will be addressed in future with more detailed physical analyses.

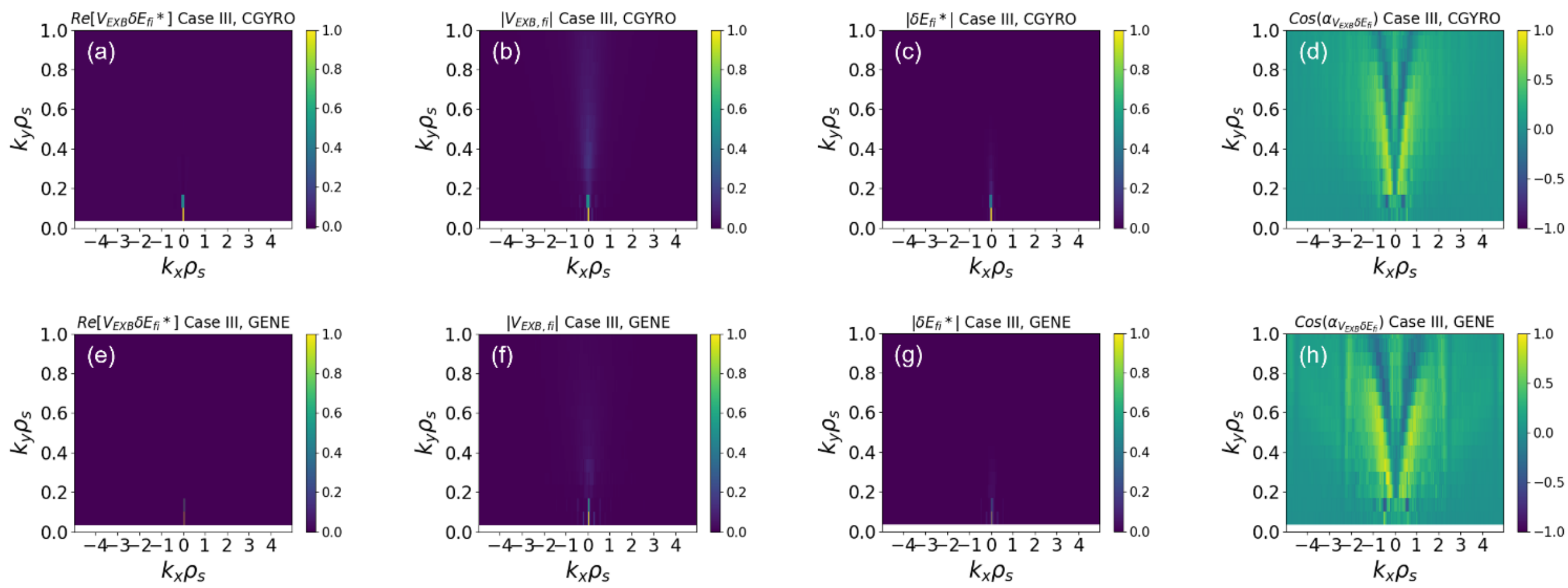

Figure 10. $\boldsymbol{k_x - k_y}$ contour of electrostatic fast ion energy flux ($\boldsymbol{Re}[\widetilde{\boldsymbol{V}}_{\boldsymbol{E\times B}}\widetilde{\boldsymbol{E}}_{\boldsymbol{f}}^{*}]$), and its fluctuating components ($\widetilde{\boldsymbol{V}}_{\boldsymbol{E\times B}}$, $\widetilde{\boldsymbol{E}}_{\boldsymbol{f}}$, and the cosine of phase angle between $\widetilde{\boldsymbol{V}}_{\boldsymbol{E\times B}}$ and $\widetilde{\boldsymbol{E}}_{\boldsymbol{f}}$) for Case III predicted by CGYRO (top) and GENE (bottom)

### *3.4 Prediction of Zonal flow level and its shearing rate*

We also examined the changes in the ratio of fluctuating electrostatic potential with zero toroidal mode number, $n$, i.e., $n = 0$ ($\delta\phi_{n=0}$), to the levels of non-zero components of the potential ($\delta\phi_{n>0}$) across four cases, as shown in Figure 11(a). This ratio can be an indicator of changes in zonal flow levels for each case. We can first notice that both codes consistently predict a reduction in this ratio with the addition of fast ions, regardless of rotation effects, i.e., Case III/Case I or Case IV/Case II. While the predicted changes in the ratio agree within their uncertainties in the cases with rotation effects (Case IV/Case II), CGYRO predicts a much smaller reduction than GENE in the cases without rotation effects (Case III/Case I). A more significant discrepancy emerges in the changes due to consideration of rotation effects. Both codes predict a reduction in the potential ratio when rotation is added without fast ions (Case II/Case I). However, this trend is significantly altered in the presence of fast ions (Case IV/Case III). While CGYRO still predicts a reduction in the potential ratio with the addition of rotation effects, GENE shows an increase in the ratio under the same conditions. In addtion, a qualitatively consistent trend across the cases discussed in this study is observed in the $n = 0$ electrostatic potential itself, not just in the ratio.

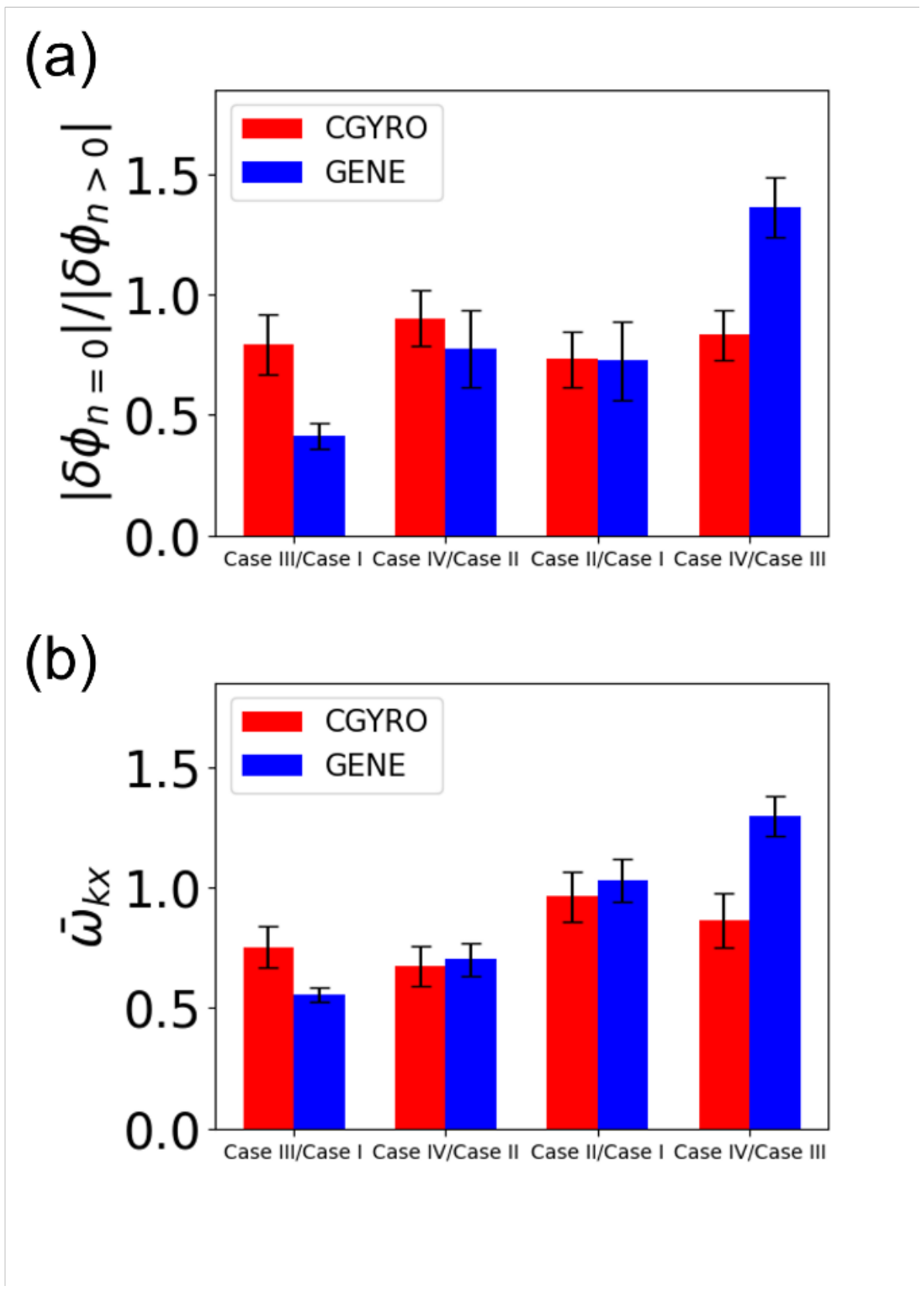

Figure 11. Changes in (a) the ratio between level of fluctuating

electrostatic potential with $\boldsymbol{n=0}$ ($\boldsymbol{\delta\phi_{n=0}}$) and the levels of $\boldsymbol{n>0}$ components of the potential ($\boldsymbol{\delta\phi_{n>0}}$), and (b) average shearing rate $\boldsymbol{\bar{\omega}_{k_x}}$ over the range of $\boldsymbol{-1.0 \leq k_x\rho_s \leq 1.0}$ where $\boldsymbol{\bar{\omega}_{k_x} = \frac{1}{N}\sum_{k_x=-\rho_s^{-1}}^{\rho_s^{-1}} \omega_{k_x}}$ and $\boldsymbol{\omega_{k_x} = \sqrt{\frac{k_x^4}{B^2}|\delta\phi\ (k_x, k_\theta = 0)|^2}}$

Since turbulence is directly suppressed by zonal shearing rather than the amplitude of zonal flow, we analyzed changes in the zonal shearing rate, $\omega_{k_x}$, defined as,

$$\omega_{k_x} = \sqrt{\frac{k_x^4}{B^2}|\delta\phi\ (k_x, k_\theta = 0)|^2} \quad (2)$$

Figure 11(b) shows the changes in the average shearing rate over the range of $-1.0 \leq k_x\rho_s \leq 1.0$, i.e., $\bar{\omega}_{k_x} = \frac{1}{N}\sum_{k_x=-\rho_s^{-1}}^{\rho_s^{-1}} \omega_{k_x}$ across the four studied. As indicated from Equation (2), the shearing rate can be considered as a $k_x^2$-weighted sum of potential. Consequently, $k_x = 0$ component does not contributed to the shearing rate. In contrast to the potential ratio comparison, we observed that the changes in the average shearing rate with the addition of fast ions were getting closer (Case III/Case I) or even within their uncertatines (Case IV/Case II) between CGYRO and GENE. This is also consistent with the agreement in the reduced energy flux ratio when fast ions are added, as discussed in Section 3.1, suggesting that the shearing rate is a more relevant parameter to energy transport than the potential ratio since the latter includes $k_x = 0$ component, which does not contribute to energy transport.

When rotation effects are added, both CGYRO and GENE predict almost no change in the cases without fast ions (Case II/Case I). However, for cases with fast ions, CGYRO predicts that the shearing rate with rotation effects slightly decrease compared to the rate without rotation effects, whereas GENE predicts an increased shearing rate under the same conditions. It is noteworthy that the energy flux levels nearly double when rotation effects are added with fast ions, i.e., from Case III to Case IV, in the GENE simulations, even with the zonal shearing rate shown in Figure 11(b). This suggests that zonal shearing effects may not be critical in determining energy flux levels in the cases investigated. Nevertheless, it is important to highlight the discrepancies in zonal flow and shearing rate preditions between two codes, as shown in Figure 11.

## 4. Conclusion

This study conducted a cross-verification of the widely-used gyrokinetic codes, CGYRO and GENE, focusing on the impact of fast ions and rotation effects on energy flux within the context of the KSTAR L-mode plasma profile. The primary aim was to characterize the similarities and differences between these codes from the perspective of users engaged in turbulence transport research and transport coefficient evaluation. Through this research, we have highlighted the importance of phase angle analysis in gyrokinetic code verification, recognizing the necessity of including phase angle analysis in the gyrokinetic analysis of fast ion effects on turbulence. Furthermore, as noted by Terry et al. [80], we concluded that it is essential to examine quantities at the lower levels of the primacy hierarchy, in addition to energy flux, since discrepancies at higher levels can manifest in divergent results at lower levels.

The similarities and differences between CGYRO and GENE observed in this study can be summarized as follows:

- Consistencies between CGYRO and GENE:
  - Linear stability analysis results
  - Fractional changes in energy flux levels and zonal shearing
  - Fast ion energy flux levels
  - Changes in fluctuations and flux distribution in k-space with fast ions
  - Changes in phase angle distribution in k-space with fast ions

- Discrepancies identified:
  - Absolute levels of thermal energy, not influenced by fast ions, requiring resolution
  - The effects of rotation on thermal transport either without or with fast ions, becoming more severe in the presence of fast ions

The qualitative consistency between CGYRO and GENE, as discussed earlier, suggests that the physical mechanisms captured by both codes align well and are similarly reproduced. While the relative transport levels appear significantly different, the absolute values remain within ~O(1) GyroBohm, which is notably smaller than the generally reported O(10)–O(100) GyroBohm order. Given this low transport regime, partly due to presence of fast ions, achieving high relative precision between the codes is inherently challenging. However, considering the small absolute discrepancies, the transport predictions of both codes can be deemed quantitatively reasonable. This implies that, for future Verification & Validation (V&V) experiments, it is advisable to first conduct experiments with larger transport magnitudes for more meaningful comparisons.

Several factors could account for the observed differences between the codes. First, differences in numerical methods used by the two codes are likely contributors. Additionally, as highlighted in prior research [21], the ExB shear is implemented differently in each code, potentially leading to varying effects. Second, while the radial box size and resolution were within similar ranges, they were not identical. Differences in initial conditions, numerical treatment of magnetic field geometry (although derived from the same gEQDSK file), and other computational factors could also contribute to the absolute discrepancies. These considerations underscore the importance of sensitivity analysis and uncertainty quantification in cross-verification, which we leave for future work. Third, practical limitations in computational resources constrained our ability to fully explore the convergence of physical quantities across numerous parameters. For instance, very low delta $k_x$ (i.e., larger radial domain) is necessitated for GENE depending on the ExB shear amplitude, requiring a convergence test. Similarly, the value for its hyperdiffusivity was set based on empirical rules, also warranting convergence testing. Moreover, convergence of transport-related quantities in nonlinear simulations as well as mode frequency and growth rate in linear simulations with respect to grid resolution is essential. The input parameters determined by convergence tests in this study demanded significant computational resources (approximately a week or more with ~O(1000) processes for a single case). Although higher resolution tests were intended, they could not be conducted due to practical constraints. From a user's perspective, when evaluating transport coefficients, it would be more practical to estimate statistical values using various codes at realistic resolutions rather than excessively increasing resolution. This highlights the importance of a diverse ecosystem of gyrokinetic codes. Lastly, as these codes continue to evolve and incorporate better methods, attention and clarification to the version used during V&V is crucial to make the reports more imformative.

Future efforts should focus on resolving the identified discrepancies, and further investigation is warranted into newly observed physical phenomena, such as the phase angle structure and its impact on energy transport for the fast ions. Based on the implications of this study, KSTAR is preparing dedicated shots for V&V, and further V&V efforts will continue.

## Acknowledgements

This research was supported by the National Research Foundation of Korea(NRF) grant funded by the Korea government(MSIT) (Grant No. RS-2022-00155991, RS-2022-00155956), the Korea Institute of Energy Technology Evaluation and Planning and the Ministry of Trade, Industry & Energy (MOTIE) of the Republic of Korea (Grant No. 20214000000410), and the R&D program of the Korea Institute of Fusion Energy (2024-EN2401-15). Computing resources were provided on the KFE computer KAIROS, funded by the Ministry of Science and ICT of the Republic of Korea (No. KFE-EN2441-10), and the National Supercomputing Center with supercomputing resources including technical support (KSC-2024-CRE-0204).